\documentclass[pdflatex,sn-vancouver,Numbered]{bst/sn-jnl}

\usepackage{graphicx}%
\usepackage{multirow}%
\usepackage{amsmath,amssymb,amsfonts}%
\usepackage{amsthm}%
\usepackage{mathrsfs}%
\usepackage[title]{appendix}%
\usepackage{xcolor}%
\usepackage{textcomp}%
\usepackage{manyfoot}%
\usepackage{booktabs}%
\usepackage{algorithm}%
\usepackage{algorithmicx}%
\usepackage{algpseudocode}%
\usepackage{listings}%
\usepackage{booktabs}
\usepackage[graphicx]{realboxes}
\usepackage{graphicx}
\usepackage{adjustbox}
\usepackage{subcaption}
\usepackage{nameref}
\usepackage{verbatim}
\usepackage{xcolor}

\theoremstyle{thmstyleone}%
\theoremstyle{thmstyletwo}%

\theoremstyle{thmstylethree}%

\newcommand{\dself}[2]{$d_{\text{self}} = #1 \pm #2$}

\newcommand{\dref}[2]{$d_{\text{ref}} = #1 \pm #2$}

\newcommand{\dselfdrefpl}[3]{$d_{\text{self}} < d_{\text{ref}},\ \text{Welch's } t(#1) = #2,\ p < #3$}

\newcommand{\totalN}{N = 1,086}   

\begin{document}

\title[Article Title]{Quantifying the Persistence of Daily Routines}


\author*[1]{\fnm{Nguyen} \sur{Luong}}\email{nguyen.luong@aalto.fi}

\author[1]{\fnm{Talayeh} \sur{Aledavood}}\email{talayeh.aledavood@aalto.fi}

\affil*[1]{\orgdiv{Department of Computer Science}, \orgname{Aalto University} \city{Espoo}, \country{Finland}}


\abstract{Daily life is structured by recurring routines that coordinate biological rhythms with social and occupational demands. Individual differences in work schedules, family obligations, and social commitments produce distinctive ways of organizing activities throughout the day. Do people have \textit{typical days} with certain arrangement of activities? How often do these typical days or routines occur and does this differ from person to person? We introduce a framework for quantifying such recurring routines, their persistence over time and their distinctiveness for different people. We model consecutive days in one's life as a sequence of different types of typical days, i.e. routines. Characterizing each day through patterns of activities common among all people -- sleep, movement, and device use -- we identify a small set of routine types that capture the dominant structure of everyday behavior. We then test whether individuals maintain stable, person-specific distributions over these types and transition between them in characteristic ways. Validating this framework with passive sensing data from 1,086 participants across 153,000 person-days in three longitudinal studies, we find that daily life typically resolves into approximately eight routine types. Each person maintains a characteristic distribution over these types, with their two most frequent routines accounting for over half of all days. Both the time allocation across routine types and the day-to-day transition dynamics are substantially more similar within individuals than between them, remaining stable across observation windows spanning weeks to months and across populations differing in age, occupation, and health status. Routine persistence shows modest associations with personality traits such as conscientiousness, but is broadly similar across age and gender. Our findings establish routine patterns as stable, person-specific behavioral fingerprints with applications in personalized health monitoring.}


\maketitle

\section{Introduction}\label{sec1}

Daily life is organized around recurring patterns of behavior — including sleep, physical activity, communication, and movement — that align biological rhythms with social, occupational, and environmental demands. People follow such patterns to adapt to work and life schedules, coordinate with others, and reduce cognitive load \cite{monk1994regularity}. Limited time, energy, and cognitive resources constrain both the range of behaviors people engage in and how predictably they repeat across days, creating a characteristic structure in daily life. Greater regularity in this daily organization predicts better health outcomes, including improved well-being, life satisfaction, and sleep quality \cite{heintzelman2019routines, margraf2016social, carney2006daily}. 

A growing body of research has shown that these behavioral patterns are not only regular, but also person-specific. Saramäki et al. first demonstrated this by showing that individuals allocate their communication effort across social contacts in a persistent, individual way, a concept they termed the social signature \cite{saramaki2014persistence}. Similar persistence has since been observed in other behavioral domains, such as mobility \cite{alessandretti2020scales}.  However, this evidence has accumulated within individual behavioral domains. It remains unknown whether persistence generalizes to routines that span multiple behaviors simultaneously. For example, a remote worker may display highly regular digital activity yet lack the mobility regularity characteristic of daily commuting. Here, we take a holistic approach by characterizing routines across multiple behavioral domains simultaneously, capturing the interplay between sleep, physical movement, and digital device use. We define an individual routine signature as (i) the proportion of days belonging to each recurring routine type and (ii) the way individuals move between these routine types over time. We hypothesize that this signature is unique to the individual and persistent over time.

The persistence of daily behavior has long been documented through surveys and diaries, which consistently reveal predictable, person-specific temporal patterns, such as meal timing and sleep schedules, and spatial regularities, such as commuting routes \cite{hansonSystematicVariabilityRepetitious1988a, monk1990social, soehnerCircadianPreferenceSleepWake2011, vagni2018patterns}. With the ubiquity of personal digital devices, everyday interactions leave behind digital traces, such as GPS coordinates, accelerometer readings, and communication logs, that can be used to quantify behavior continuously and unobtrusively over long periods. Studies leveraging such data have consistently found strong persistence within individual behavioral domains, including sleep, mobility \cite{songLimitsPredictabilityHuman2010, song2010modelling, alessandretti2020scales}, and communication \cite{saramaki2014persistence, aledavoodDailyRhythmsMobile2015a, aledavood2016channel, heydari2018multichannel, loriteLongTermEvolutionEmail2016, urena2026multiplexity}, as well as digital behaviors such as web visits \cite{barbosa2016returners, hu2018life, kulshrestha2021web} and application usage \cite{malmi2016you, kosinski2013private, petersSocialMediaUse2024, shawBehavioralConsistencyDigital2022}. However, persistence within a single behavior does not necessarily imply that a coherent routine emerges when multiple behaviors are considered together. It is therefore necessary to examine whether persistence generalizes to routines that span multiple behaviors simultaneously.

Because daily life is constrained by limited time and recurring demands, people tend to settle into a limited set of recurring daily patterns across days. They return to a limited set of locations \cite{alessandrettiUnderstandingInterplaySocial2018} and maintain a stable circle of close contacts \cite{saramaki2014persistence}. Likewise, when multiple behaviors are considered together, individuals tend to cycle through a small number of recurring routines. For a college student, this might include a routine of going to campus at specific times and attending classes, alongside a distinct routine of staying home and relaxing. Such routines can be represented as a small set of interpretable types that capture the most common ways people organize their days \cite{farrahiDiscoveringRoutinesLargescale2011, vagni2018patterns}. Eagle et al. introduced the related concept of eigenbehaviors \cite{eagleEigenbehaviorsIdentifyingStructure2009a}, showing that a small number of underlying dimensions, or principal components, can explain much of the variation in daily life, each corresponding to an interpretable routine such as ``morning class,'' ``evening social,'' or ``weekend home.'' Similar structures have since been identified in mobility \cite{jiangClusteringDailyPatterns2012, farrahiDiscoveringRoutinesLargescale2011}, activity behaviors \cite{yangIdentifyingLatentActivity2023a}, and daily behaviors \cite{aledavood2022quantifying, girardini2023adaptation, farrahiWhatDidYou2008}.

This work has two main objectives. First, we test whether the persistence observed in single behavioral domains generalizes to routines spanning multiple behaviors simultaneously. Second, we assess whether this persistence holds across populations that differ in life circumstances and demographics. To address the first objective, we ask whether individuals show stable, person-specific distributions over routine types over time. We do this by identifying a small set of recurring routine types from patterns of daily behavior. We find that people tend to settle into a characteristic set of routine types, forming a personal routine signature with greater similarity within individuals than between individuals. To address the second objective, we replicate the analysis across three independent studies spanning populations that differ in age, occupation, and life circumstances, and find that the routine signature is robust across all three studies, holding over both short (weeks) and long (months) observation windows.


\section*{Results}\label{sec:results}  

\subsection*{Daily life resolves into a small number of recurring routine types}

We analyzed three independent longitudinal datasets spanning \totalN{} participants and over 153,000 person-days in total. The Tesserae study (N=592) followed information workers across the United States over one year. The MoMo-Mood study (N=164) tracked patients with depressive episodes and healthy controls in Finland over one year. The GLOBEM study (N=497) collected passive sensing data from university students in the United States across four academic years. Despite the substantial differences between the populations, daily behaviors consistently resolved into a compact set of recurring routine types when using clustering methods on the data. Each day was characterized by measures of sleep, physical movement, and digital device use, summarized across morning, afternoon, evening, and night segments. Eight routine types sufficed to capture the dominant structure of daily life in each study (see Methods).

\begin{figure}[htbp!]
    \centering
    \includegraphics[width=1\linewidth]{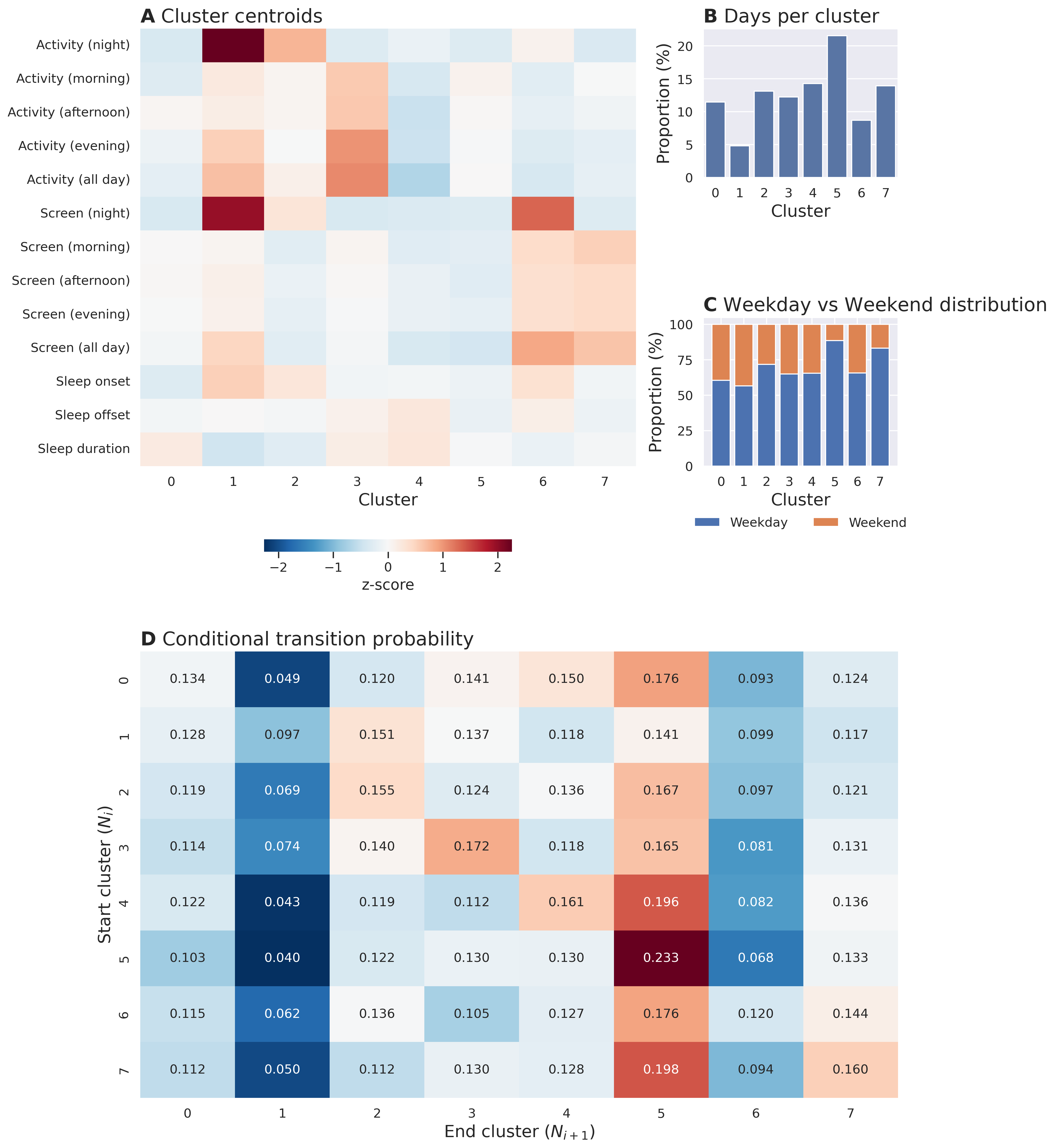}
    \caption{Cluster summary for Tesserae. (A) Cluster centroids: Heatmap showing the average standardized feature values (z-scores) for each behavioral feature within each cluster. Each feature’s z-score is computed per individual to reflect within-person deviations. Higher (red) and lower (blue) values represent positive and negative deviations from an individual’s mean, respectively. (B) Days per cluster: Histogram showing the distribution of all recorded days across clusters, aggregated across the study population. (C) Weekday–weekend distribution: Stacked bar plot depicting the proportion of weekdays (blue) and weekends (orange) within each cluster. Certain clusters exhibit a significant weekend skew, depicting leisure-oriented behaviours that emerge on non-working days. (D) Heatmap of adjacency matrix depicting the conditional transition probability between cluster pairs. Transition probabilities are computed and normalized per participant, then averaged across participants to obtain a population mean. Higher values indicate more frequent transitions.}
    \label{fig:tesserae-cluster-summary}
\end{figure}

The routine types were interpretable and reflected the everyday structure of life in each population. \autoref{fig:tesserae-cluster-summary}A shows the cluster centroids for the Tesserae study, expressed as within-person deviations from each individual's own baseline. One dominant routine type captured the most typical pattern of daily life in this predominantly white-collar population, with most features near their individual means and slightly reduced smartphone screen use (\autoref{fig:tesserae-cluster-summary}B). Other routine types appeared less frequently but reflected recognizable alternatives, including patterns resembling weekends or holidays, such as slightly longer sleep and higher smartphone screen use (Cluster 4) or increased nocturnal mobility (Cluster 6) (\autoref{fig:tesserae-cluster-summary}C). Notably, these patterns emerged without providing calendar information to the clustering algorithm, indicating that they simply arose from behavioral regularities in the data. Corresponding routine types for MoMo-Mood and GLOBEM are shown in Appendix B.

Having identified the routine types, we next examined how individuals transition between them from day to day. If individuals were confined to a single routine, transitions would be dominated by self-transitions, and the transition matrix would be concentrated along the diagonal. However, this was not the case. As shown in \autoref{fig:tesserae-cluster-summary}D, the average transition matrix for Tesserae exhibits diffuse transitions rather than a strong diagonal structure. Several routine types tend to transition toward a common dominant type, rather than returning to themselves. The same pattern was observed in MoMo-Mood and GLOBEM (SI Appendix), indicating that individuals do not adhere to a single routine but instead cycle through multiple routines in a structured, non-random manner.

\subsection*{Individuals maintain a distinctive and stable routine signature}\label{sec:results:signature_persistence}

The routine types identified above describe daily life at the population level as they are the recurring patterns that emerge when we pool days from everyone. Individuals, however, differ in how they allocate their time across these types: one person may spend most of their days in one routine, while another divides their time more evenly across several. We next test whether this distribution of time across routine types is distinctive between people and persistent within a person over time.

To test this, we computed a routine signature for each participant: the proportion of their days spent in each routine type, ordered from most to least frequent (see \autoref{sec:methods}). We then split each participant's observation period into two equal halves — 135 days each for Tesserae and MoMo-Mood, 30 days each for GLOBEM — and computed two quantities for every person. The self-distance measures how much a person's signature in the second half of their data differs from their signature in the first half, capturing within-person persistence. The reference distance measures how much that same person's signature differs from the signatures of other participants observed over the same window, capturing between-person distinctiveness. Intuitively, a signature should produce small self-distances and larger reference distances — in other words, a person should look more like their own self than like anyone else. We quantified both distances using the Jensen–Shannon divergence and performed a robustness check using cosine distance.

\begin{figure}
    \centering
    \includegraphics[width=1\linewidth]{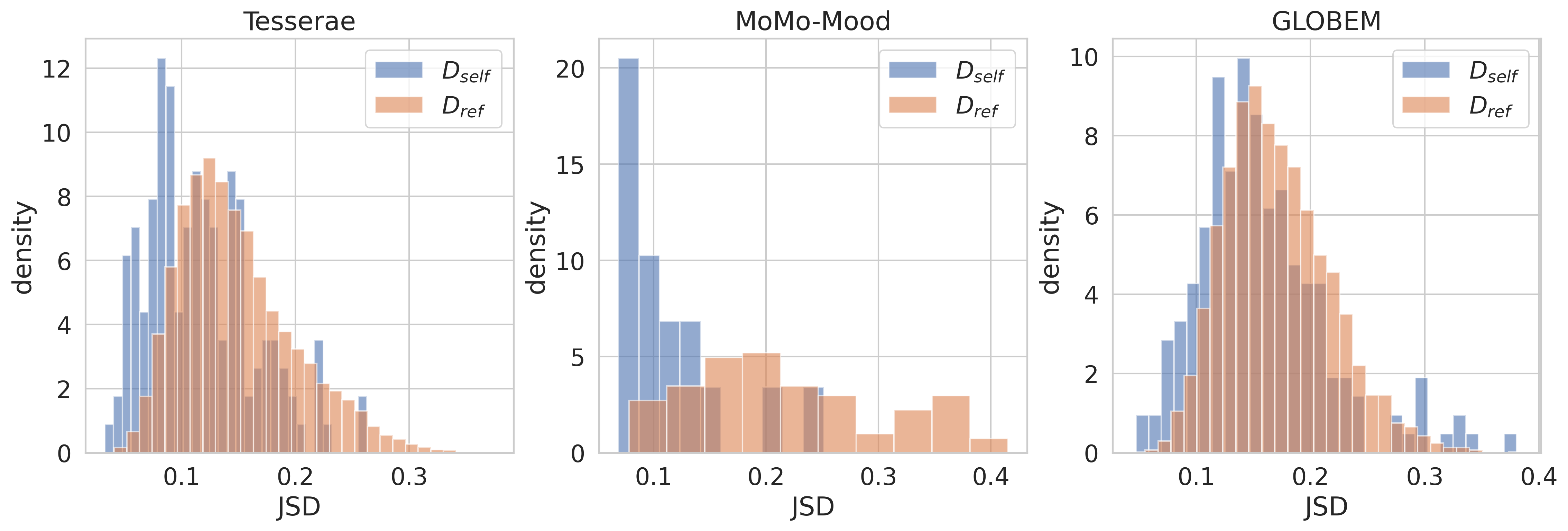}
    \caption{Self-distance and reference-distance of routine signature across studies.}
    \label{fig:dself_dref}
\end{figure}

Two findings emerged. First, despite the apparent variety of everyday life, people concentrated most of their time in only a handful of routine types.
Across all three studies, the two most frequent routine types accounted for $(57.63 \pm 11.43)\%$ of a participant's days on average, meaning more than half
of daily life was spent in just two recurring patterns, with the remaining six types filling in the rest. Second, each person's routine signature was
substantially more similar to their own than to the signatures of other individuals measured over the same time window (\autoref{fig:dself_dref}). In Tesserae, within-person self-distances were consistently smaller than between-person reference distances, both for Jensen-Shannon divergence
(\dself{0.116}{0.047} vs.\ \dref{0.148}{0.023};
\dselfdrefpl{213.87}{-7.3}{10^{-3}}) and cosine distance
(\dself{0.028}{0.027} vs.\ \dref{0.052}{0.016};
\dselfdrefpl{240.26}{-9.38}{10^{-3}}). The same pattern held in MoMo-Mood
and GLOBEM. Crucially, this
persistence was visible both over short observation windows of a few weeks in GLOBEM and over long windows of several months in Tesserae and MoMo-Mood,
indicating that routine signatures are not a short-term artifact of a particular season.

\subsection*{Transition between routines exhibits similar persistence}

The routine signature captures how a person allocates days across routine types but does not give information about the order in which those days occur. Two people could spend identical fractions of their time in the same routines and yet live very different weeks: one regularly alternates between a weekday and a weekend pattern while the other moves irregularly through several routine types. We therefore asked whether the way a person moves between routine types from one day to the next is itself person-specific and stable over time.

To test this hypothesis, we constructed a transition probability matrix $\mathbf{P} \in \mathbb{R}^{K \times K}$ for each participant and time window (see \autoref{sec:methods:signature}), and computed self-distance and reference-distance analogously to the routine signature, using the mean row-wise Jensen-Shannon divergence between matrices.

\begin{figure}
    \centering
    \includegraphics[width=1\linewidth]{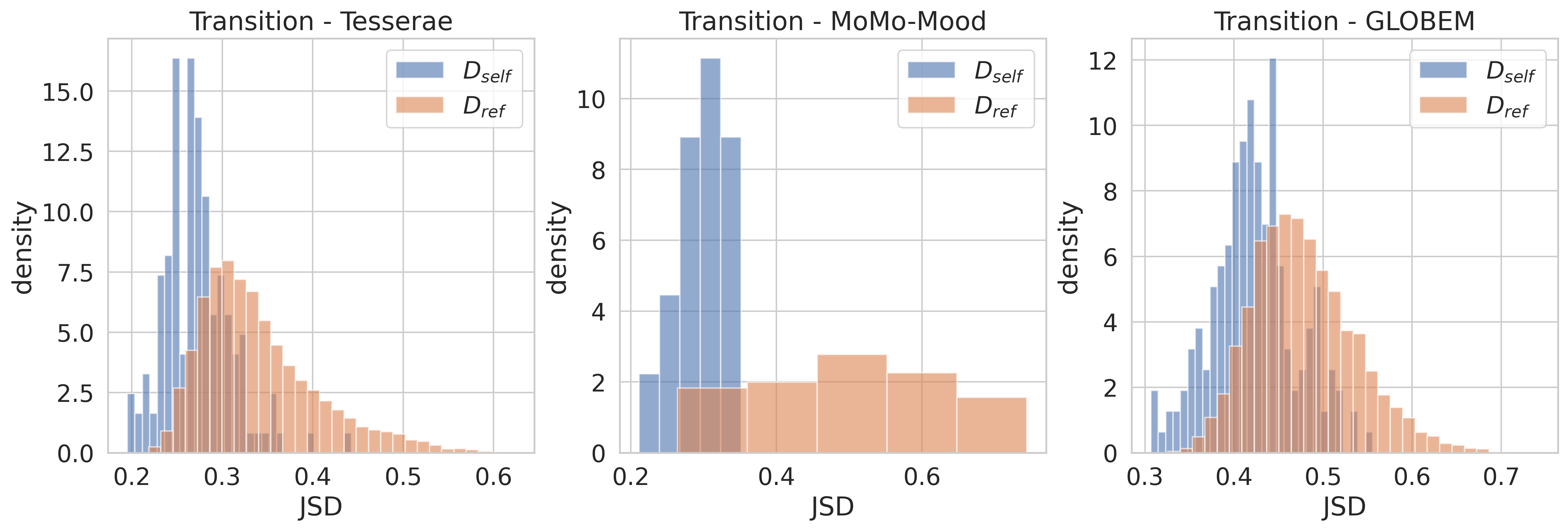}
    \caption{Self-distance and reference-distance of transition signature across studies.}
    \label{fig:transition-signature}
\end{figure}

\autoref{fig:transition-signature} presents the distributions of within- and between-person distances.
Across all three cohorts, transition signatures were temporally stable and individually distinctive, with within-person distances consistently smaller than between-person distances. The effect was clearest in Tesserae, where a large sample of information
workers was followed over two 135-day windows: \dself{0.273}{0.038} compared with \dref{0.341}{0.040}
(\dselfdrefpl{292.19}{-14.96}{10^{-3}}). The same pattern held, with a larger gap between within- and between-person distances, in MoMo-Mood, despite a smaller clinical cohort observed over the same
long windows: \dself{0.297}{0.034} compared with \dref{0.524}{0.058}
(\dselfdrefpl{22.34}{-13.17}{10^{-3}}). In GLOBEM, where each participant was observed over much shorter 30-day windows, the gap was narrower but
still robust: \dself{0.565}{0.077} compared with \dref{0.598}{0.025}
(\dselfdrefpl{238.37}{-5.59}{10^{-3}}).

\subsection*{Factors predicting persistence of routine signature}\label{sec3.3}

Although the results presented above suggest that routines are typically stable, that stability is not uniform across individuals. Prior work has linked this variability to demographic factors and personality traits, primarily in the context of
single-behavior routines such as communication or mobility \cite{kulshrestha2021web, alessandrettiUnderstandingInterplaySocial2018, centellegherPersonalityTraitsEgonetwork2017}. We asked whether the same predictors relate to the stability of routine signatures. Using each participant's within-person self-distance as the outcome - with smaller values indicating more stable routines - we fit linear regression models with age, gender, and Big Five personality traits as predictors. Because Tesserae and MoMo-Mood share the same long observation windows, we pooled them into a single long-window analysis. GLOBEM, with its shorter 30-day windows and repeated measurements across academic years, was analyzed separately using a linear mixed-effects model with participant-level random intercepts (Methods). We fit parallel models for the transition signature distance ($d_{\text{self}}^{\text{trans}}$). 

The results of long-window analysis computed on Tesserae and MoMo-Mood are shown in \autoref{tab:ols_signature_long}.
\begin{table}[htbp]
\centering
\caption{OLS results for long-term $d_{\text{self}}$ and $d_{\text{self}}^{\text{trans}}$.}
\label{tab:ols_signature_long}
\begin{tabular}{lcccccc}
\toprule
& \multicolumn{3}{c}{$d_{\text{self}}$} & \multicolumn{3}{c}{$d_{\text{self}}^{\text{trans}}$} \\
\cmidrule(lr){2-4}\cmidrule(lr){5-7}
Predictors & Estimate & 95\% CI & $p$ & Estimate & 95\% CI & $p$ \\
\midrule
(Intercept)            & 0.139   & $0.050$--$0.228$    & \textbf{0.003} & 0.233    & $0.159$--$0.307$    & \textbf{$<0.001$} \\
Extraversion           & $-0.004$& $-0.014$--$0.007$   & 0.501          & 0.001    & $-0.007$--$0.010$   & 0.738 \\
Agreeableness          & $-0.013$& $-0.027$--$0.000$   & 0.057          & $-0.007$ & $-0.018$--$0.005$   & 0.241 \\
Conscientiousness      & 0.015   & $0.004$--$0.026$    & \textbf{0.006} & 0.006    & $-0.002$--$0.015$   & 0.155 \\
Neuroticism            & $-0.005$& $-0.016$--$0.006$   & 0.362          & 0.005    & $-0.004$--$0.014$   & 0.259 \\
Openness               & 0.008   & $-0.004$--$0.020$   & 0.196          & 0.007    & $-0.003$--$0.017$   & 0.167 \\
Gender [Male]          & $-0.014$& $-0.031$--$0.003$   & 0.103          & $-0.002$ & $-0.016$--$0.012$   & 0.774 \\
Age bin [$\ge 25$]     & $-0.029$& $-0.054$--$-0.004$  & \textbf{0.024} & 0.001    & $-0.020$--$0.022$   & 0.920 \\
\midrule
$R^2$ / adj.\ $R^2$    & \multicolumn{3}{c}{0.107 / 0.066} & \multicolumn{3}{c}{0.040 / -0.003} \\
\bottomrule
\end{tabular}
\end{table}
Across all models, gender was not associated with the stability of either signature. In the long-window analysis, older participants (age $\geq 25$) showed lower routine signature distance ($b = -0.029$, 95\% CI $[-0.054, -0.004]$, $p = 0.024$), indicating more stable routines, though no comparable effect emerged for the transition signature. While the overall effects of personality traits were small, conscientiousness was positively linked to higher routine signature distance ($b = 0.015$, 95\% CI $[0.004, 0.026]$, $p = 0.006$). Extraversion, agreeableness, neuroticism, and openness were not significant predictors in any model. In the short-window GLOBEM analysis, no demographic or personality variables were significantly associated with either signature.

\section*{Discussion}\label{sec4}  

This work builds on prior evidence of routine persistence, examining how it manifests across different behavioral domains and populations. We introduce a framework that models routines as latent behavioral classes and quantifies persistence via each individual’s time allocation across these classes. Across three independent studies, individuals consistently allocated similar proportions of time to a small set of routines, yielding distinctive routine signatures. Beyond time allocation, day-to-day transition dynamics between routines were also person-specific, forming a complementary transition signature. Both signatures were distinctive and separable across individuals. Persistence was broadly similar across demographic groups (age, gender), but showed associations with personality traits such as conscientiousness.

To contextualize our findings, we draw comparisons with previous research on behavioral stability. Research on the social signature has demonstrated that individuals persistently allocate communication efforts unevenly across a small set of close ties, with this pattern remaining stable over time and generalizing across populations \cite{saramaki2014persistence,heydari2018multichannel, iniguezUniversalPatternsEgocentric2023}. Persistence also emerges in mobility patterns \cite{alessandrettiUnderstandingInterplaySocial2018, alessandretti2020scales} and online behaviors \cite{kulshrestha2021web, malmi2016you}, where individuals follow consistent, repeatable paths through physical and digital environments. Our work extends these domain-specific findings by demonstrating that both allocation and transition patterns exhibit similar stability across multiple behaviors, suggesting that routine persistence reflects a general principle of human behavior rather than isolated phenomena.

Our approach proceeds from a simple premise: complex behaviors can be captured by a small set of latent routine classes \cite{eagleEigenbehaviorsIdentifyingStructure2009a,aledavood2022quantifying,zhou2022predicting}. Although the approach is unsupervised, the principal findings are robust to the number of components and distance metrics: people spend most of their time in a few dominant routines, and the resulting signatures remain stable over time. This mirrors regularities in other domains: people repeatedly visit a limited set of places \cite{alessandretti2020scales} and use familiar apps \cite{sekaraTemporalCulturalLimits2021}. Importantly, this stability does not imply rigidity. For example, prior work has shown that when individuals face external constraints, such as the COVID-19 pandemic lockdowns \cite{luong2023impact, luong2024sleep, girardini2023adaptation}, they adapt in person-specific rather than uniform ways, suggesting that routine signatures reflect a characteristic that persists even as specific behaviors adjust to changing circumstances. 

However, the distinctiveness of these signatures implies re-identification risk. Prior work has demonstrated that mobility patterns \cite{alessandretti2020scales}, communication behaviors \cite{saramaki2014persistence}, and app usage \cite{sekaraTemporalCulturalLimits2021, malmi2016you} can identify individuals, but combining multiple behavioral modalities creates even more unique patterns. The temporal stability we observe means signatures could link the same individual across datasets or time periods \cite{sekaraTemporalCulturalLimits2021}, posing re-identification risk even in de-identified data. This raises a practical question: how much data is required for reliable discrimination, and does this threshold vary systematically across populations?  From a mental health perspective, a stable
routine signature may serve as an individual-specific baseline against which behavioral disruptions can be detected. This creates the possibility for personalized passive monitoring that identifies meaningful deviations from a person’s own characteristic
pattern, rather than from group-level norms \cite{huckvale2019toward, mohr2017personal}.

Our study has limitations in both the data and the methodological approach. On the data side, although we considered multiple routine modalities, common routines like location and communication were not included due to missingness and sparsity of features. GPS signals showed substantial missingness across datasets, so we excluded location features to avoid biased estimates. Communication activity was logged only via calls and SMS. Without app-based messaging, the communication feature space was sparse and zero-inflated. As a result, clustering on these variables tended to produce a dominant “no communication” cluster that mainly separated participants with any call/SMS activity from those with none. On the methodology side, our clustering approach is also limited in many ways. While the number of latent classes was selected via model selection, the model remains unsupervised in nature, meaning that the resulting clusters can vary between populations or carry different meanings, making interpretation hard. Furthermore, by clustering via a bag-of-days method, we ignored the sequential temporal order of routines. The routine signature should therefore be interpreted as a description of what a person's routine days look like and how often each kind of day occurs, rather than as a description of how those days follow one another over time. The transition signature adds some of this missing information by capturing day-to-day shifts, but it does not describe longer patterns such as weekly cycles or gradual change. Future work can consider temporal dependencies via appropriate methods such as the Markov model \cite{liuIntraindividualPhenotypingDepression2024, leaningUncoveringSocialStates2025}.

\section*{Methods}\label{sec:methods}  

\subsection*{Study description}\label{sec:methods:study_desc}  

To examine how daily routines persist across populations and sensing modalities, we analyzed three multisensor, longitudinal datasets—Tesserae (information workers), MoMo-Mood (patients with depressive episodes and healthy controls), and GLOBEM (college students). These studies combined passively collected smartphone traces with wearable data over months to years. 

\textbf{Tesserae.} The Tesserae study recruited 757 information workers across the United States \cite{mattingly2019tesserae}. The study aims to design and implement a large-scale, long-term multimodal sensing system to capture real-world behaviors, psychological states, and workplace performance among information workers. Participants were provided a Garmin Vivosmart 3 wristband to monitor mobility, heart rate, sleep, and calories. Additionally, participants installed a custom smartphone app that passively tracked screen activity, data usage, charging behavior, location, and other phone states. The study was granted IRB approval from University of Notre Dame under protocol number 17-05-3870.

\textbf{MoMo-Mood.} The MoMo-Mood study involved patients diagnosed with major depressive episodes, including Major Depressive Disorder (MDD), Bipolar Disorder (BD), Borderline Personality Disorder (BPD), as well as healthy controls (HC). The study allows us to test whether routine signatures remain person-specific and stable in a clinical population, where day-to-day behavior is more likely to be disrupted. The study recruited a total of 164 participants from Finland: 133 patients and 31 HCs \cite{aledavood2025multimodal}. Data were collected over a one-year period from smartphones using the Niima data collection platform and AWARE app \cite{aledavood2017data, ferreiraAWAREMobileContext2015}, which tracked screen activity, calls, and SMS, charging behavior, location, and accelerometer. The study was granted research permits by Helsinki and Uusimaa Hospital District Psychiatry with approval number § 125/2018.

\textbf{GLOBEM.} The GLOBEM study procured a multi-year passive sensing dataset of over 700 user-years of data from 497 unique college students \cite{xu2022globem}. The study aims to support research on developing, testing, and evaluating machine learning methods to better understand the daily behaviors, health, and well-being of college students. Longitudinal data, including location, phone usage, calls, Bluetooth, mobility, and sleep behavior, were collected from a mobile app and fitness trackers. The study spanned four years, with data collection occurring over a 10-week period each year, primarily during the spring semester. The study received IRB approval from the University of Washington with the IRB number STUDY00003244.

\subsection*{Features extraction and Preprocessing} \label{sec:methods:features_extraction}  

\begin{figure}[!htbp]
    \centering
    \includegraphics[width=\textwidth]{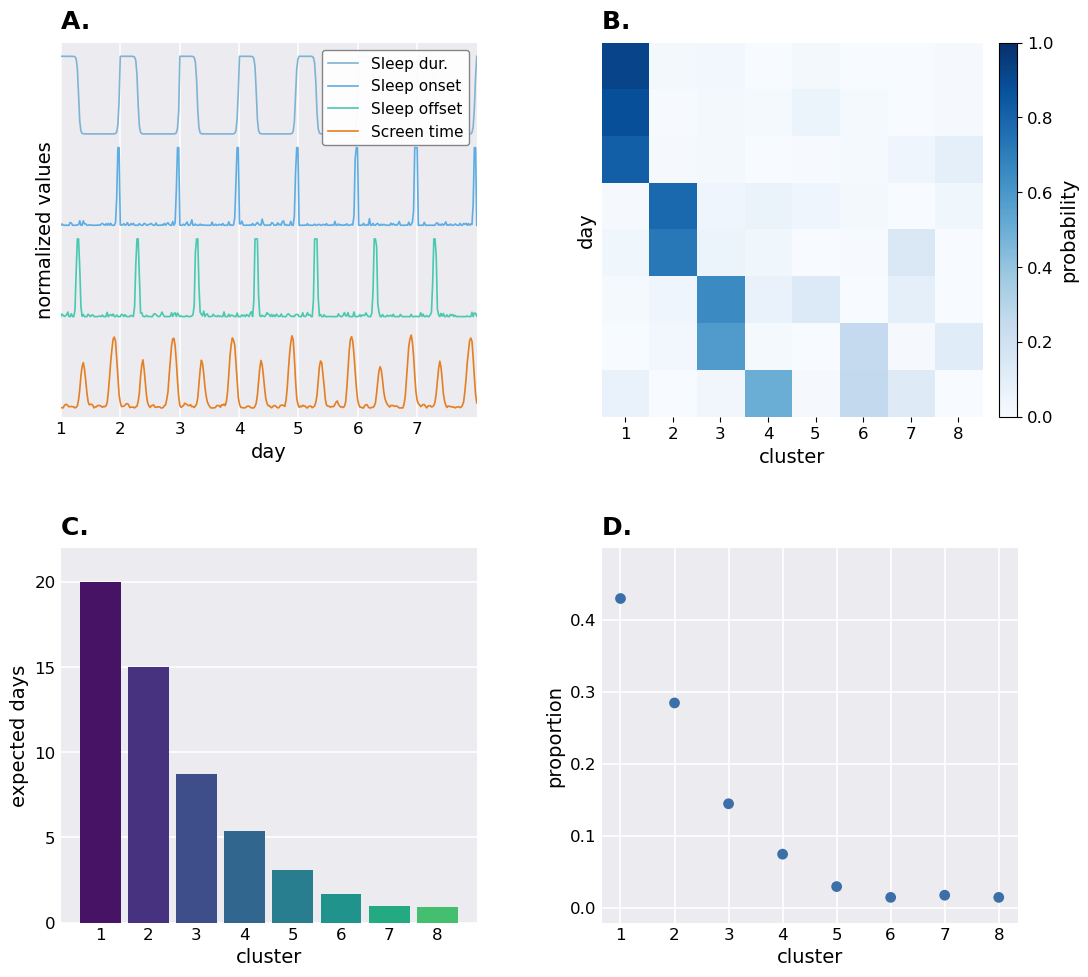}
    \caption{Schematic description of the routine signature. (A) Raw data capturing each person-day's behavior (e.g., sleep times, screen use, activity levels). (B) A Gaussian Mixture Model fitted on the pooled person-days assigns each day a vector of cluster assignment probabilities across $K$ latent routine clusters. (C) For each individual, the expected number of days in each cluster is computed by summing cluster assignment probabilities across days, and clusters are ranked in descending order of frequency. (D) These expected counts are normalized into proportions, yielding the individual's routine signature.}
    \label{fig:routine-sig-workflow}
\end{figure}

From raw data, we extracted behavioral features related to three behavioral domains: sleep, mobility, and device usage. Demographics and qualitative measures of personality traits were gathered from the baseline surveys. An overview of feature processing and building of routine signature is presented in \autoref{fig:routine-sig-workflow}.
 
\textbf{Sleep.} Bedtime, wake time, and sleep duration were collected directly from Garmin (Tesserae) and Fitbit (GLOBEM) fitness trackers. In MoMo-Mood, due to the lack of wearable devices, sleep was estimated using phone lock/unlock status, with the longest lock episode considered sleep. To assess the validity of this proxy, sleep parameters from the same participants were compared to those derived from actigraphy during a two-week validation phase. The comparison revealed a slight bias but demonstrated adequate correlation between the two methods.

\textbf{Mobility.} This routine was measured using step count data collected from fitness trackers in the Tesserae and GLOBEM studies. In MoMo-Mood, due to the lack of fitness trackers, mobility was estimated using the standard deviation of the accelerometer magnitude \cite{ravi2005activity}. Accelerometer readings displayed a clear 24-hour rhythm, consisting of higher activity levels during the day, a peak around a specific hour, and a gradual decline toward the end of the day. To capture daily activity patterns, activity readings were aggregated into four 6-hour time bins as a proxy for four parts of the day (night, morning, afternoon, evening). A daily average value was also computed.

\textbf{Device usage.} The hourly duration of screen use episodes was extracted. Similar to mobility, daily rhythms for screen usage were constructed by aggregating the hourly measures into 6-hour bins. 

\textbf{Demographics and personality traits.} Baseline demographics (age, gender, occupation) were recorded at enrollment. In Tesserae, age was available only in aggregated bins; to ensure comparability, we applied the same binning across all studies: \(<\!25\), \(25\text{--}34\), \(35\text{--}44\), \(45\text{--}54\), and \(55\text{--}64\). Personality traits were assessed with the BFI-10 questionnaire \cite{rammstedt2007measuring} in Tesserae and GLOBEM and the NEO-60 questionnaire \cite{costa1992neo} in MoMo-Mood. Because Tesserae reported personality traits as aggregate scores on a 1–5 scale, we linearly rescaled the MoMo-Mood and GLOBEM trait scores to the same $[1,5]$ range for cross-study comparability.

\textbf{Exclusion and normalization}: The records from the first and last day of any user were excluded due to the high likelihood of data incompleteness. We excluded participants with fewer than 14 days of data for the clustering task. Any day with missing data on device usage, mobility, or sleep was excluded. A total of N=1,086 participants were retained after this filtering process. Each feature was z-normalized within participants, allowing the model to focus on intra-individual variability. The resulting daily vector contained 13 features (see \autoref{tab:features_list}). We did not apply dimensionality reduction because the feature set is small, and keeping the original variables preserves the interpretability of the clustering results.

\begin{table}[ht]
    \centering
    \begin{tabular}{ll}
        \toprule
        \textbf{Routine} & \textbf{Features} \\
        \midrule
        Mobility & Step count (MoMo-Mood, GLOBEM); accelerometer (Tesserae): daily average + MAEN \\
        Sleep             & bedtime, wake time, sleep duration \\
        Device usage      & Screen usage duration: daily average + MAEN \\
        \bottomrule
    \end{tabular}
    \caption{List of daily behavioral features, grouped by routine domain. MAEN = Morning, Afternoon, Evening, Night segments.}
    \label{tab:features_list}
\end{table}

\subsection*{Building routine clusters using GMM}\label{sec:methods:building_cluster}
We pooled daily features across all participants in a study into a single $D \times N$ matrix ($N$ features over $D$ person-days) and fitted one $K$-component GMM at the population level. The $K$ multivariate Gaussian components (means $\boldsymbol{\mu}_k$, covariances $\boldsymbol{\Sigma}_k$, weights $\pi_k$) represent latent routines shared across the population. This approach captures both the mean behavioral profile (via the components' means) and the variability and correlation among features (via the covariance matrices). Compared to clustering methods like K-Means, GMM offers greater flexibility for modelling the fluid nature of daily behavior. First, GMM supports soft clustering: rather than committing each day to a single routine, the model assigns each day a vector of membership probabilities $(u_{d1}, \ldots, u_{dK})$ with $\sum_k u_{dk} = 1$. This is appropriate for modelling daily life, where transitional days such as a sick day that disrupts a typical weekday, rarely conform to a single routine. Second, each GMM component can be parameterized by a full covariance structure, which captures inter-dependencies between features. For instance, increased phone activity late at night may correlate with a delayed bedtime.

Tuning GMM involves two main hyperparameters: the number of components \(k\) and the covariance type (full, tied, diagonal, or spherical). Model selection was based on the Bayesian Information Criterion (BIC), with lower values indicating a better fit. For the model with a full covariance matrix, we quantified cluster separation by computing all pair-wise Bhattacharyya distances. For two multivariate Gaussian clusters \(p_1 \sim \mathcal{N}(\boldsymbol{\mu}_1, \boldsymbol{\Sigma}_1)\) and \(p_2 \sim \mathcal{N}(\boldsymbol{\mu}_2, \boldsymbol{\Sigma}_2)\), the Bhattacharyya distance is

\[
D_B(p_1,p_2)
    = \frac{1}{8}\,
      (\boldsymbol{\mu}_1 - \boldsymbol{\mu}_2)^{\mathsf T}
      \boldsymbol{\Sigma}^{-1}
      (\boldsymbol{\mu}_1 - \boldsymbol{\mu}_2)
      + \frac{1}{2}\,
        \ln\!\left(
          \frac{\det\boldsymbol{\Sigma}}
               {\sqrt{\det\boldsymbol{\Sigma}_1\,\det\boldsymbol{\Sigma}_2}}
        \right),
\]

where
\[
\boldsymbol{\Sigma} = \frac{1}{2} \left( \boldsymbol{\Sigma}_1 + \boldsymbol{\Sigma}_2 \right)
\]
is the average covariance matrix, and \(\boldsymbol{\mu}_1,\boldsymbol{\mu}_2\) are the cluster mean vectors. Larger values indicate better separation between clusters.

We retained the full vector for each day rather than collapsing to a single label. This clustering process was carried out independently for each study. For GLOBEM, the dataset spanned multiple academic years, so we fitted the clustering model separately for each year (semester). The tuning result is presented in Appendix A.

\subsection*{Routine signature}\label{sec:methods:signature} 

Social signature describes how individuals persistently distribute their communication efforts across their social contacts \cite{saramaki2014persistence}. We extended this concept by introducing the notion of a routine signature: the characteristic way in which individuals distribute their time across recurring patterns of daily behavior. Our concept aims to capture the persistence with which individuals structure their daily lives across different types of routines.

For each participant, we represented every day by its vector of cluster assignment probabilities from the GMM $\mathbf{u}_d^{(i)} = (u_{d1}^{(i)}, \ldots, u_{dK}^{(i)})$. Then, we summarized how often each cluster occurred for that participant by summing the cluster assignment probabilities across all of their days, yielding the expected number of days $w_{ij} = \sum_d u_{dj}^{(i)}$ spent in each cluster $j$. To obtain a comparable distribution across individuals, these expected counts were normalized into proportions, giving the fraction of the participant's days that belonged to each cluster.

The clusters were then ranked in descending order of frequency (rank 1 = the most common cluster for that participant). The resulting vector of ranked proportions constitutes the \textit{routine signature} of that individual:

\begin{equation}
    \sigma_i = \left(
    \frac{w_{i1}}{\sum_j w_{ij}},\;
    \frac{w_{i2}}{\sum_j w_{ij}},\;
    \ldots,\;
    \frac{w_{ik_i}}{\sum_j w_{ij}}
    \right),
\end{equation}

where $w_{ij} = \sum_d u_{dj}^{(i)}$ is the expected number of days participant $i$ spent in cluster $j$, and the elements are ordered so that $\sigma_{i1} \geq \sigma_{i2} \geq \dots \geq \sigma_{ik}$ with $k$ being the number of clusters.

To additionally capture the order in which routines occur, we define a \textit{transition signature} based on day-to-day movement between clusters. For each participant, we computed a $K \times K$ matrix whose entry $T_{jk}^{(i)}$ is the expected number of $j \to k$ transitions, obtained by summing the outer products of consecutive days' cluster assignment probabilities:

\begin{equation}
    T_{jk}^{(i)} = \sum_{d=1}^{M_i - 1} u_{dj}^{(i)} \cdot u_{d+1,k}^{(i)}.
\end{equation}

Row-normalizing $T^{(i)}$ yields the conditional transition probability $P_{jk}^{(i)} = T_{jk}^{(i)} / \sum_{k'} T_{jk'}^{(i)}$, which gives the probability of moving to cluster $k$ given the current cluster $j$. The resulting matrix $\mathbf{P}^{(i)}$ constitutes the transition signature of participant $i$.

Intuitively, the routine signature describes what routines a person inhabits, while the transition signature describes how they move between them. In line with the concept of social signatures, we expect both patterns to persist over time.

\subsection*{Persistence of routine signatures}\label{sec:methods:signature_persistence}  

To quantify the persistence of routine signatures, we examine how robust a person maintains the time allocated to their routine over time. The persistence of routine was defined as the distance between the routine signatures in consecutive time segments, using the Jensen-Shannon divergence (JSD) \cite{lin1991divergence}:

\begin{equation}
JSD(\sigma_1, \sigma_2) = H(\frac{\sigma_1 + \sigma_2}{2}) - \frac{1}{2}[H(\sigma_1) + H(\sigma_2)]
\end{equation}

where $\sigma_1$ and $\sigma_2$ are the routine signatures defined in Eq. 1, and $H(\sigma)$ is the Shannon entropy of $\sigma$.

To quantify how well an individual maintains their routine over time, we define the self-distance as the JSD between their signatures in the two consecutive time splits:
\begin{equation}
d_{\text{self}}(i) = \text{JSD}\!\left(\sigma_i^{(1)}, \sigma_i^{(2)}\right),
\end{equation}
where $\sigma_i^{(1)}$ and $\sigma_i^{(2)}$ denote the routine signatures of individual $i$ in the first and second time split, respectively. To benchmark this within-person persistence against between-person variability, we compute the reference distance as the average JSD between $i$'s signatures and those of every other individual $j$, matched by time split:
\begin{equation}
d_{\text{ref}}(i) = \frac{1}{N-1} \sum_{j \neq i} \frac{1}{2}\left[\text{JSD}\!\left(\sigma_i^{(1)}, \sigma_j^{(1)}\right) + \text{JSD}\!\left(\sigma_i^{(2)}, \sigma_j^{(2)}\right)\right],
\end{equation}
where $N$ is the number of individuals in the study. Comparing signatures within the same time split controls for any temporal shifts in the population (e.g., seasonal effects) that could otherwise inflate between-person distances. If $d_{\text{self}}(i) < d_{\text{ref}}(i)$, individual $i$'s routine is more similar to their own past than to others' routines, indicating a persistent and person-specific signature.

\newpage

\subsection*{Declarations}

\subsubsection*{Availability of data and material}

The GLOBEM dataset is publicly available through PhysioNet at \url{https://physionet.org/content/globem/}. The Tesserae dataset is made available after submission of a project / research proposal and completion of a DUA with data being restricted to non-profit entities.

Due to the highly sensitive and private nature, the data collected in the MoMo-Mood study are not publicly archived or available. Accessing the data could potentially be possible through collaboration requests.

\subsubsection*{Competing interests}

The authors declare that they have no competing interests.

\subsubsection*{Funding}

Not applicable.

\subsubsection*{Authors' contributions}

NL and TA conceived, designed, and developed the study. NL analyzed the empirical data and wrote the paper. All authors read, revised, and approved the final manuscript.

\subsubsection*{Acknowledgements}

The authors acknowledge the computational resources provided by the Aalto Science-IT project.

\bibliography{sn-bibliography}


\clearpage

\section{Appendices}

\begin{appendices}

\clearpage                    
\section{GMM model tuning}

Model tuning across all cohorts favored Gaussian mixtures with full covariance matrices, yielding lower Bayesian Information Criterion (BIC; lower is better) and higher average Bhattacharyya distances between components (greater separation). Performance improved up to \(K=8\) and then showed diminishing returns, so we fixed \(K=8\) for the main analyses. All principal findings were confirmed in a sensitivity analysis with \(K \in \{6,\ldots,11\}\).

\begin{figure}[p]
  \centering
  \includegraphics[width=\linewidth]{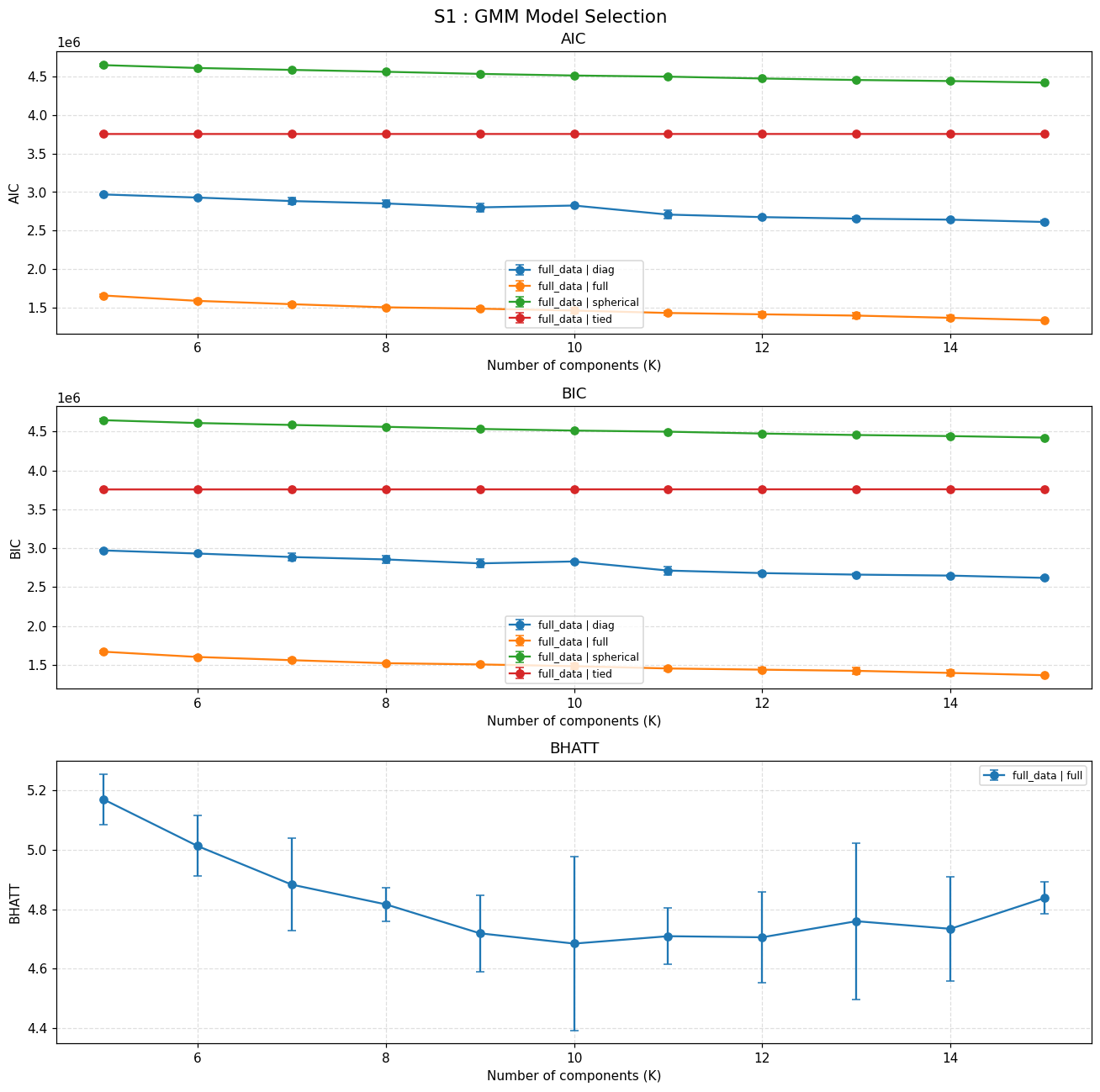}
  \caption{Tesserae: Model selection}
  \label{fig:tesserae_model_selection}
\end{figure}

\begin{figure}[p]
  \centering
  \includegraphics[width=\linewidth]{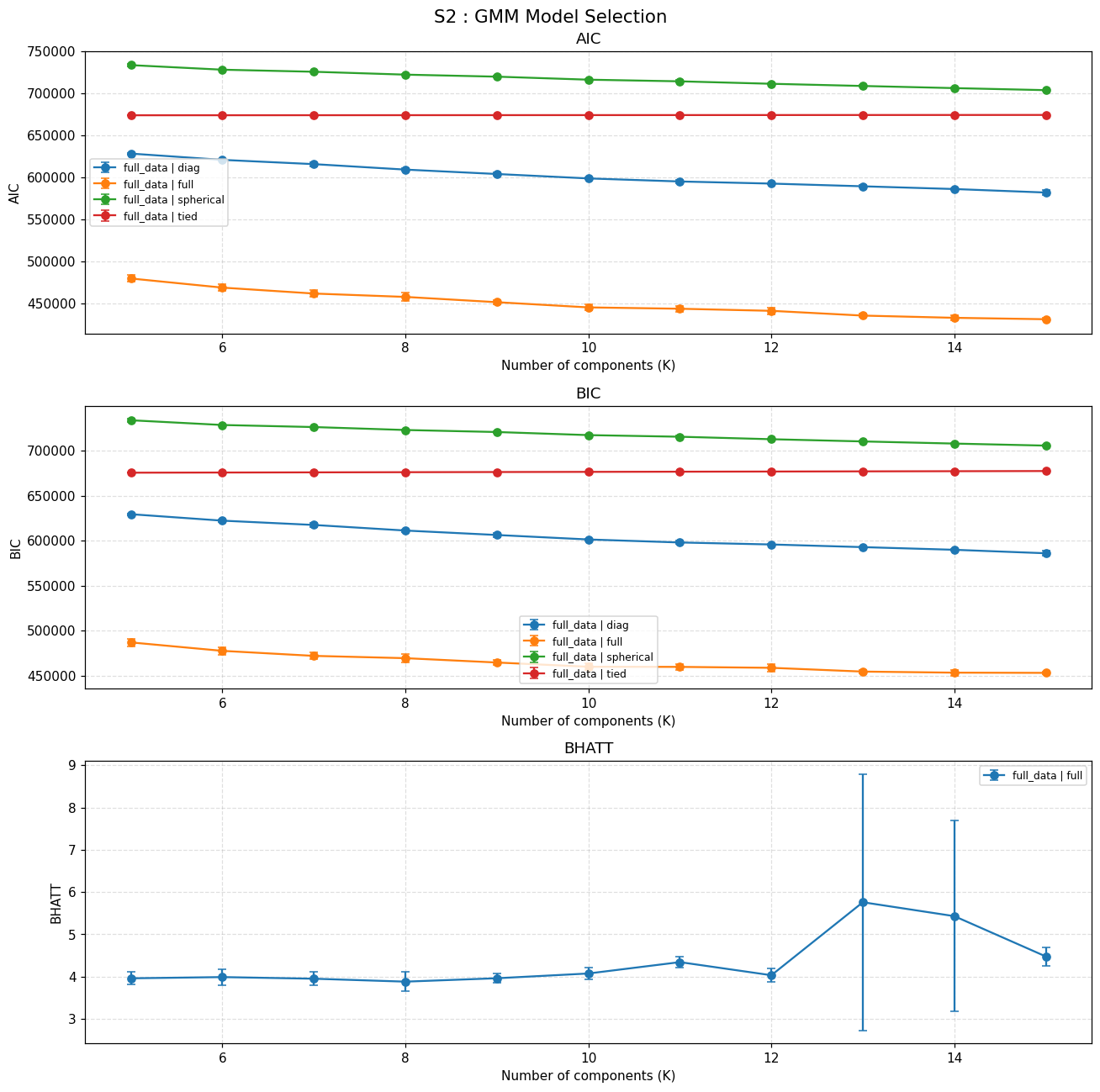}
  \caption{MoMo-Mood: Model selection}
  \label{fig:momo_gmm_model_selection}
\end{figure}

\begin{figure}[p]
  \centering
  \includegraphics[width=\linewidth]{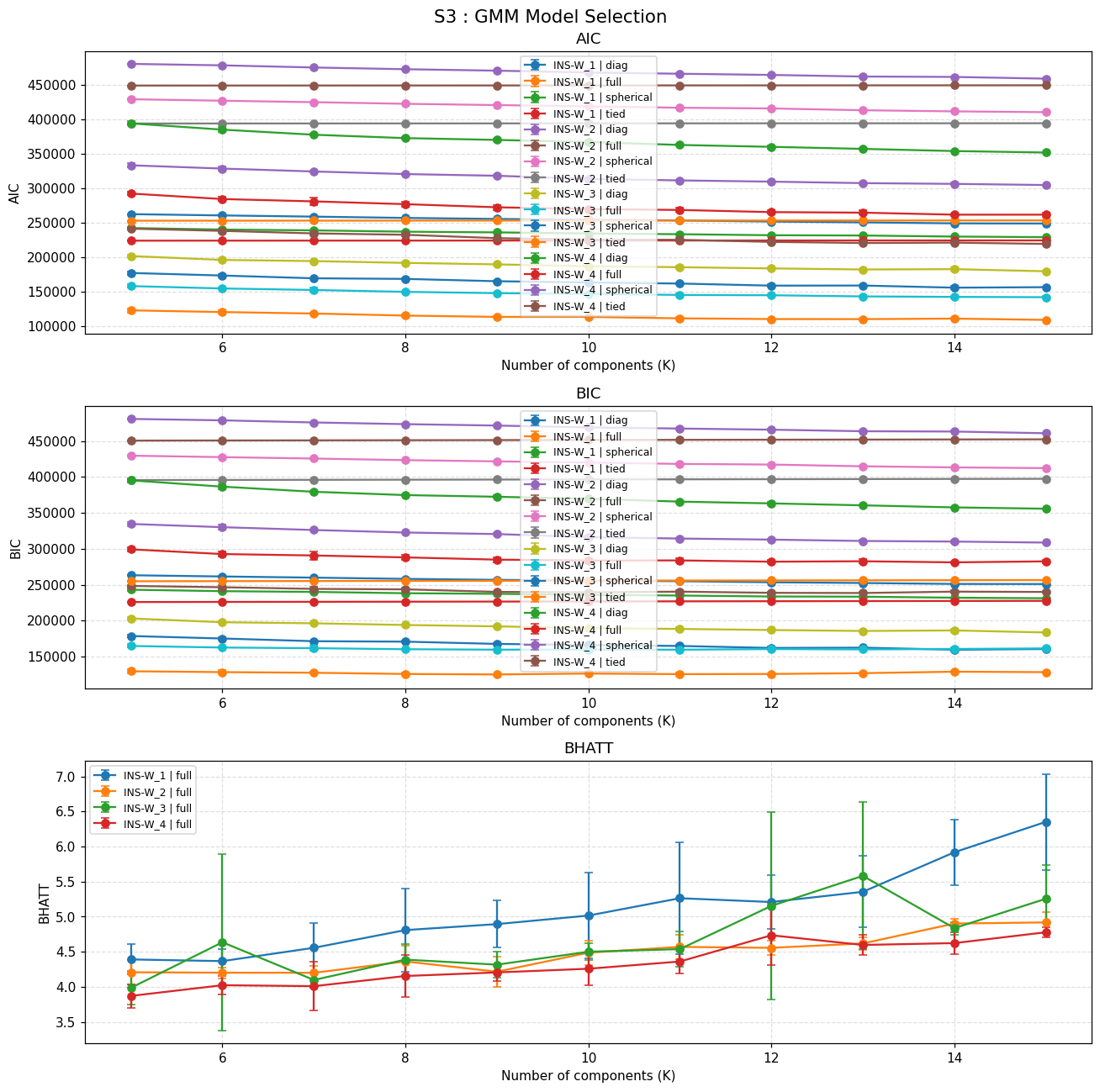}
  \caption{GLOBEM: Model selection}
  \label{fig:globem_gmm_model_selection}
\end{figure}

\clearpage
\section{Cluster properties of MoMo-Mood and GLOBEM}

\begin{figure}[p]
  \centering
  \includegraphics[width=\linewidth]{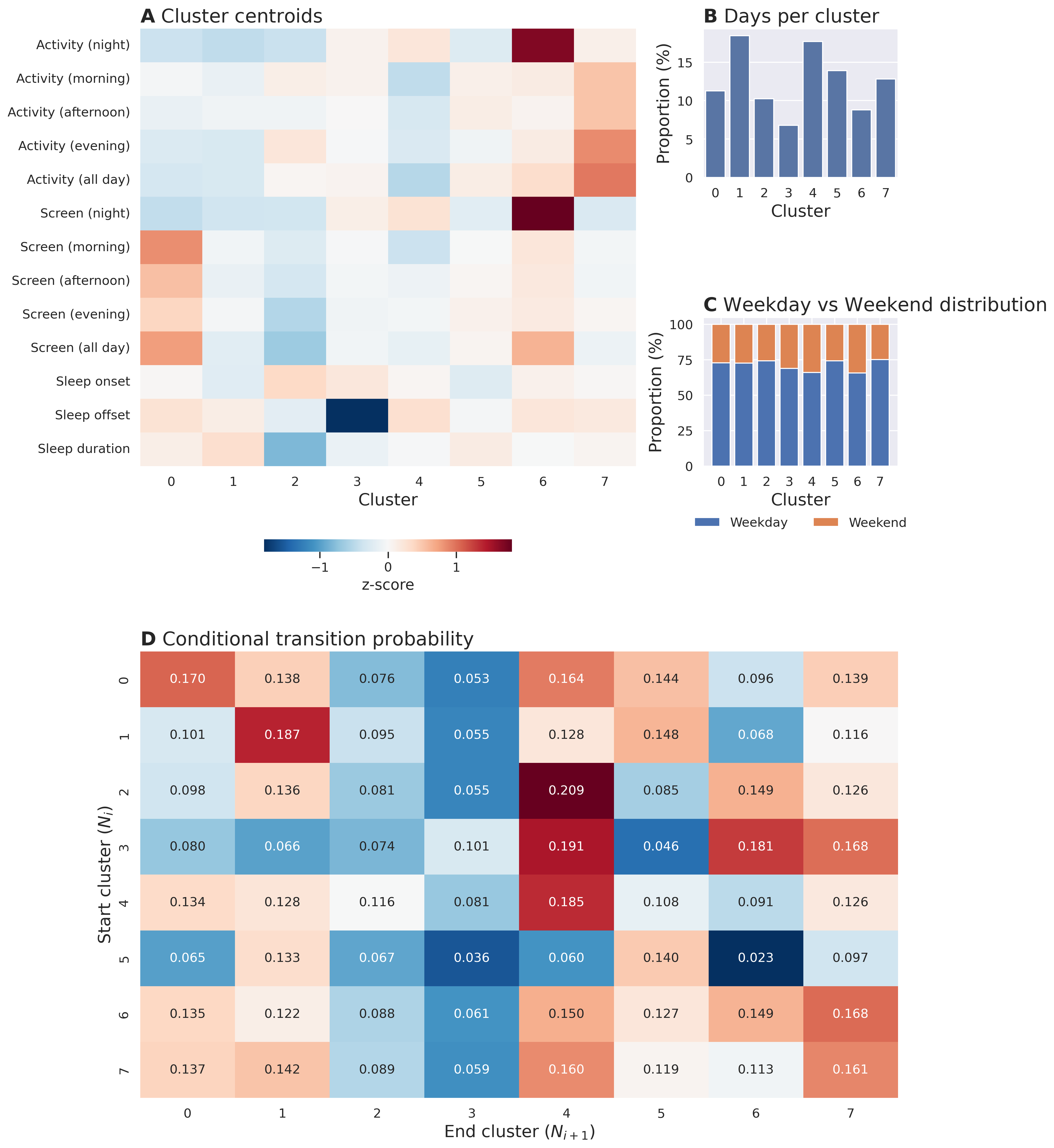}
  \caption{MoMo-Mood: Cluster centroid characteristics}
  \label{fig:momo_centroid_summary}
\end{figure}

\begin{figure}[p]
  \centering

  \begin{subfigure}[t]{0.485\textwidth}
    \centering
    \includegraphics[width=\linewidth]{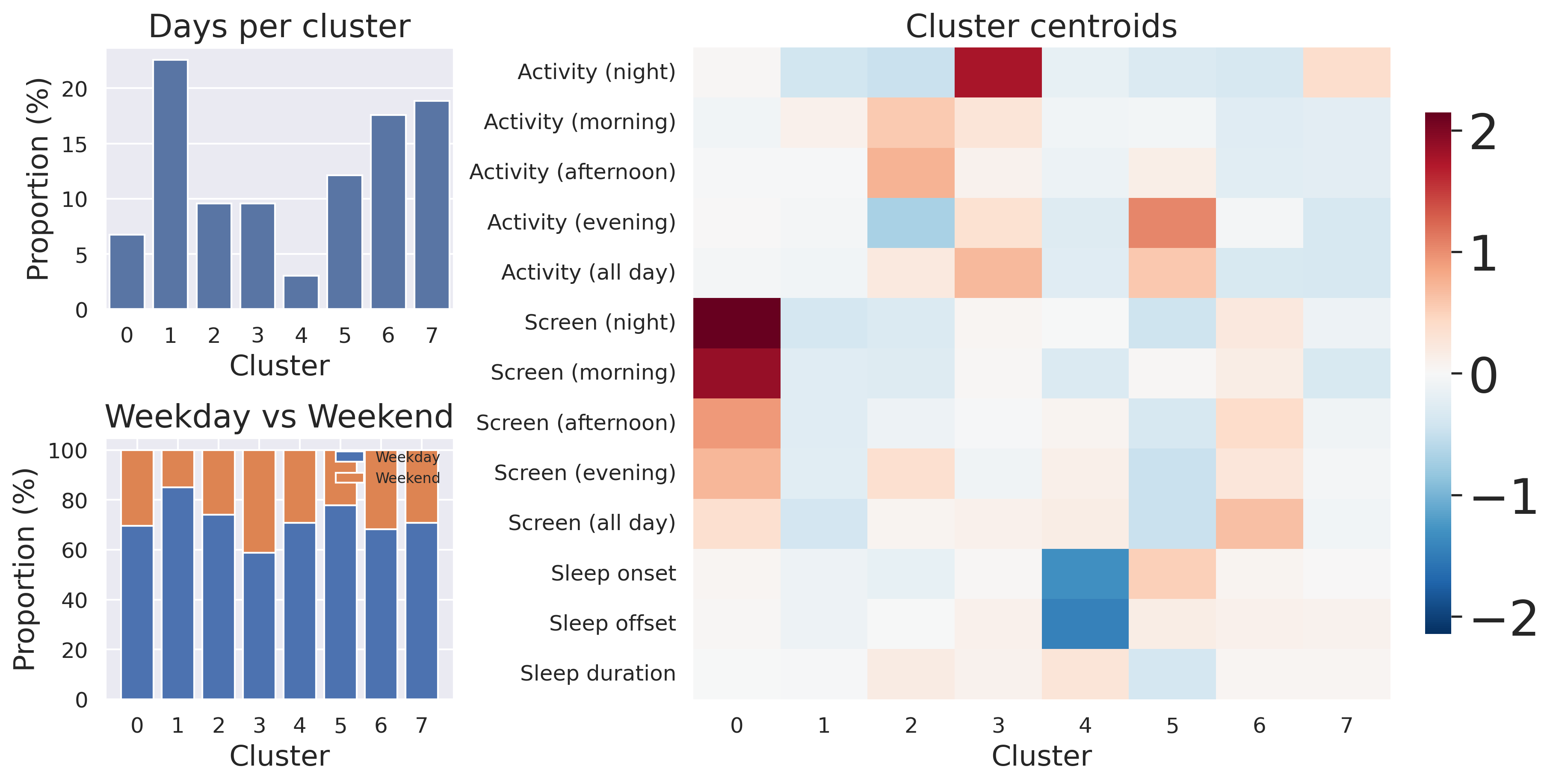}
    \caption{Cluster 1}
    \label{fig:globem_centroids_a}
  \end{subfigure}\hfill
  \begin{subfigure}[t]{0.485\textwidth}
    \centering
    \includegraphics[width=\linewidth]{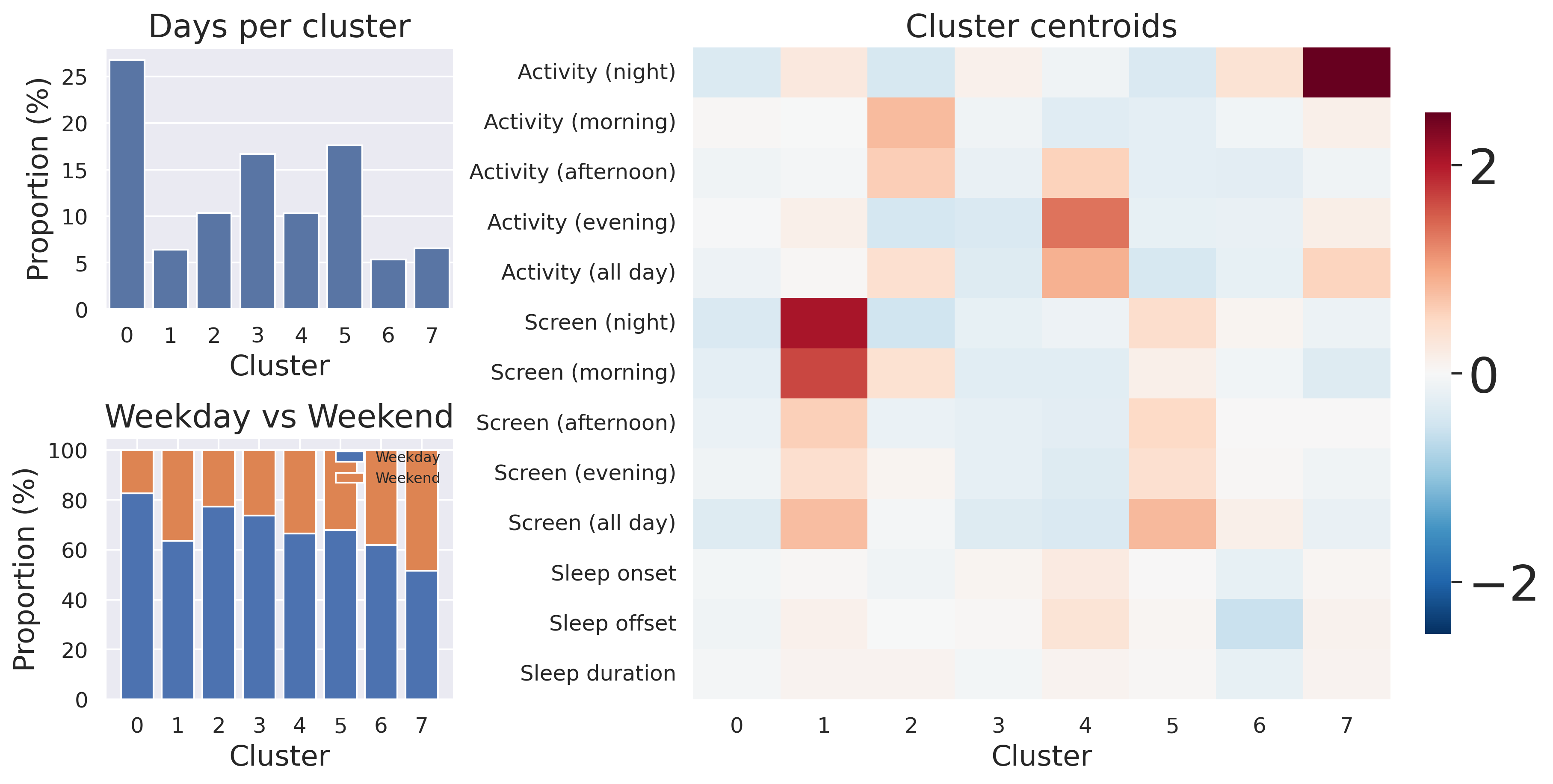}
    \caption{Cluster 2}
    \label{fig:globem_centroids_b}
  \end{subfigure}

  \medskip

  \begin{subfigure}[t]{0.485\textwidth}
    \centering
    \includegraphics[width=\linewidth]{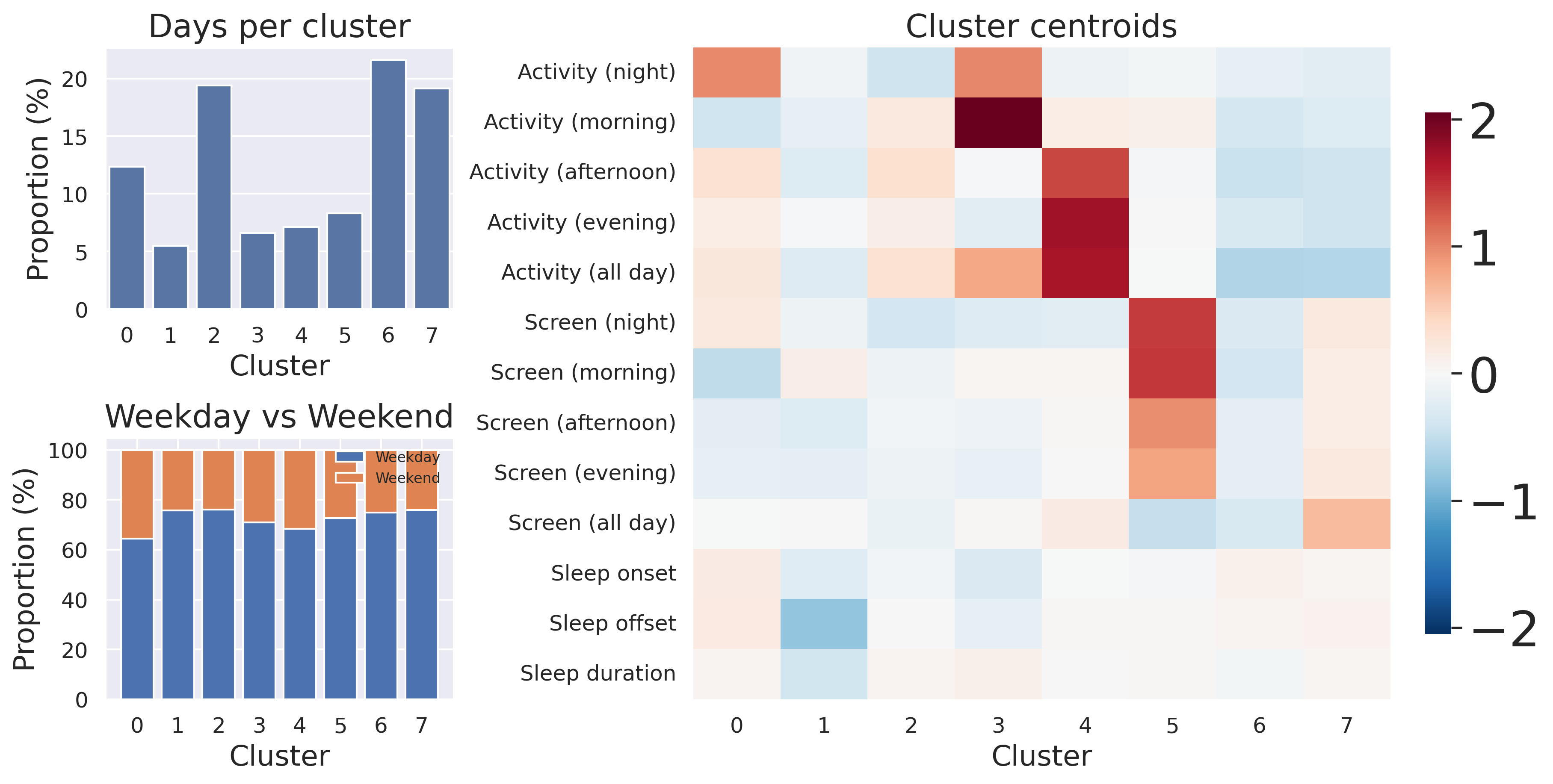}
    \caption{Cluster 3}
    \label{fig:globem_centroids_c}
  \end{subfigure}\hfill
  \begin{subfigure}[t]{0.485\textwidth}
    \centering
    \includegraphics[width=\linewidth]{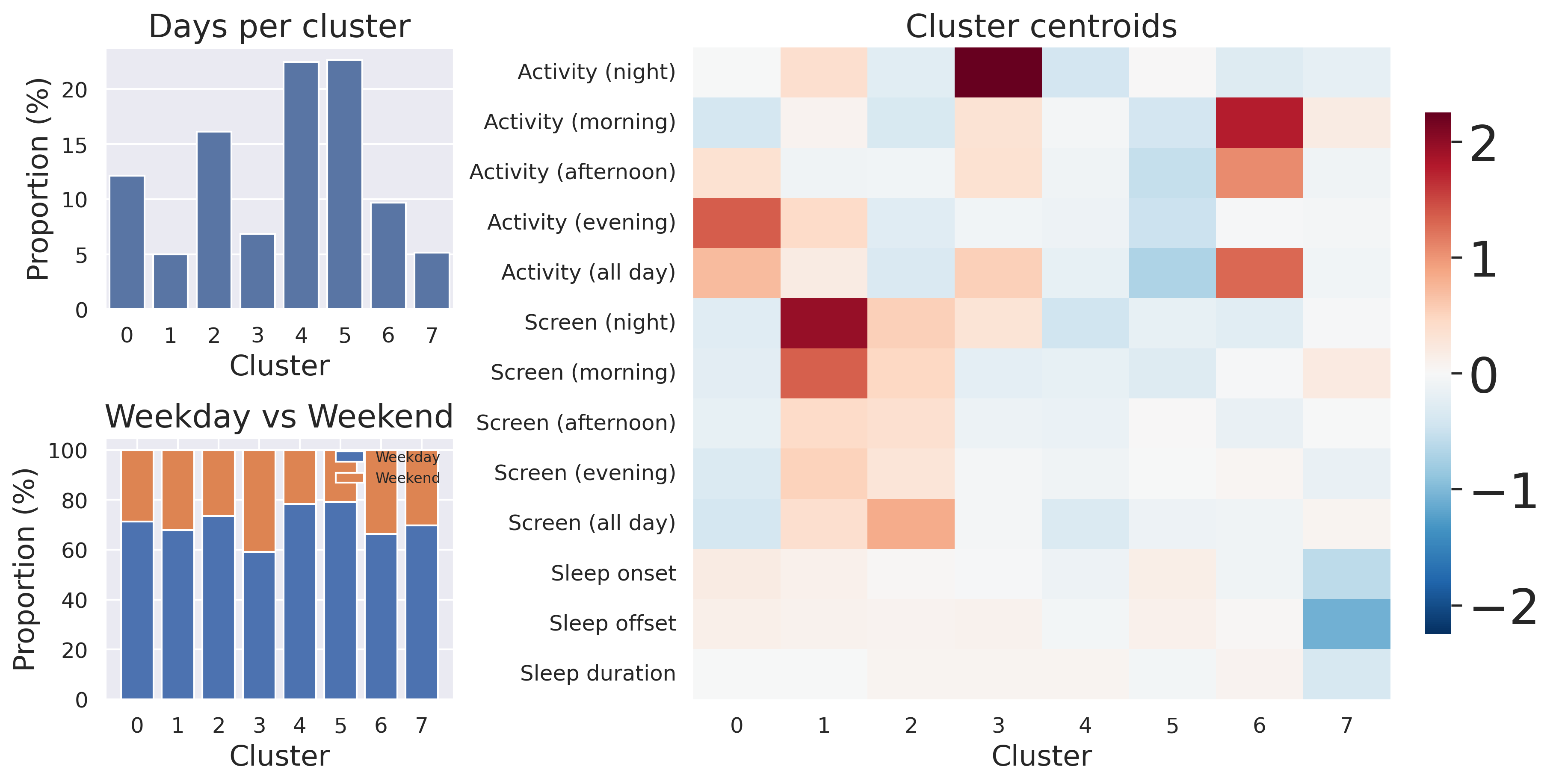}
    \caption{Cluster 4}
    \label{fig:globem_centroids_d}
  \end{subfigure}

  \caption{GLOBEM: Cluster centroid characteristics.}
  \label{fig:globem_centroids}
\end{figure}

\autoref{fig:momo_centroid_summary} and \autoref{fig:globem_centroids} describe the cluster characteristics of the MoMo-Mood and GLOBEM studies, respectively. Across both datasets, a few dominant clusters account for the majority of time, while other clusters reflect free day routines, for example, elevated nightly activity or increased nighttime screen use.

\clearpage
\section{Sensitivity check: Routine signature with varying K components}

Across all studies, rank-ordered routine proportions show the same pattern: a few dominant routines at the head (rank 1–2) and a long, light tail. Shaded bands (±1 SD across participants) are widest at rank 1 and shrink with rank, implying heterogeneity among individuals at the top routines. 

\begin{figure}[htbp]
  \centering
  \includegraphics[width=\linewidth]{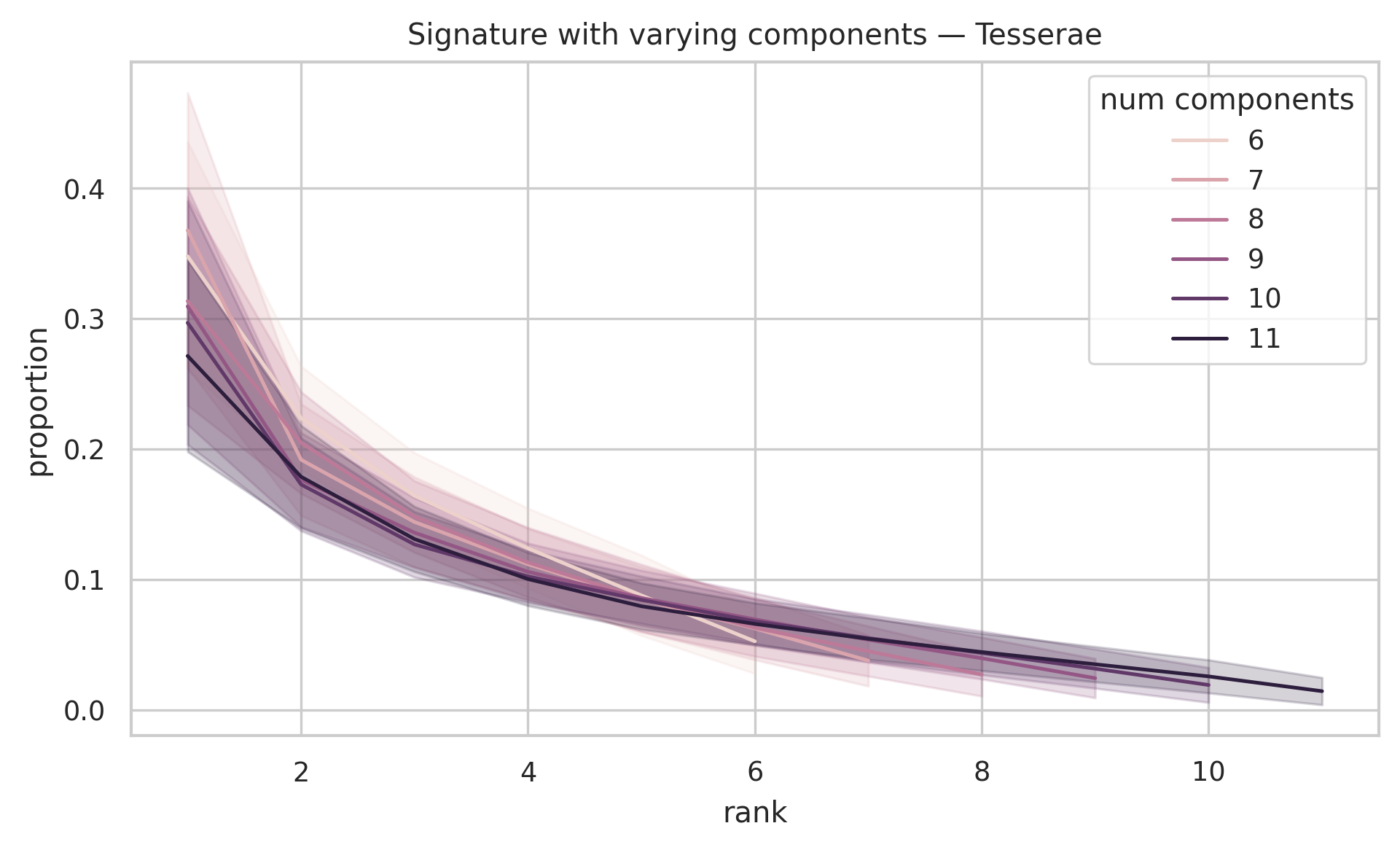}
  \caption{Tesserae. Mean proportion by rank for \(K=6\)–\(11\). A small number of routines dominate; extra components mainly split the low-rank tail. Bands show \(\pm 1\) SD across participants.}
  \label{fig:signature_varyingK_tesserae}
\end{figure}

\begin{figure}[htbp]
  \centering
  \includegraphics[width=\linewidth]{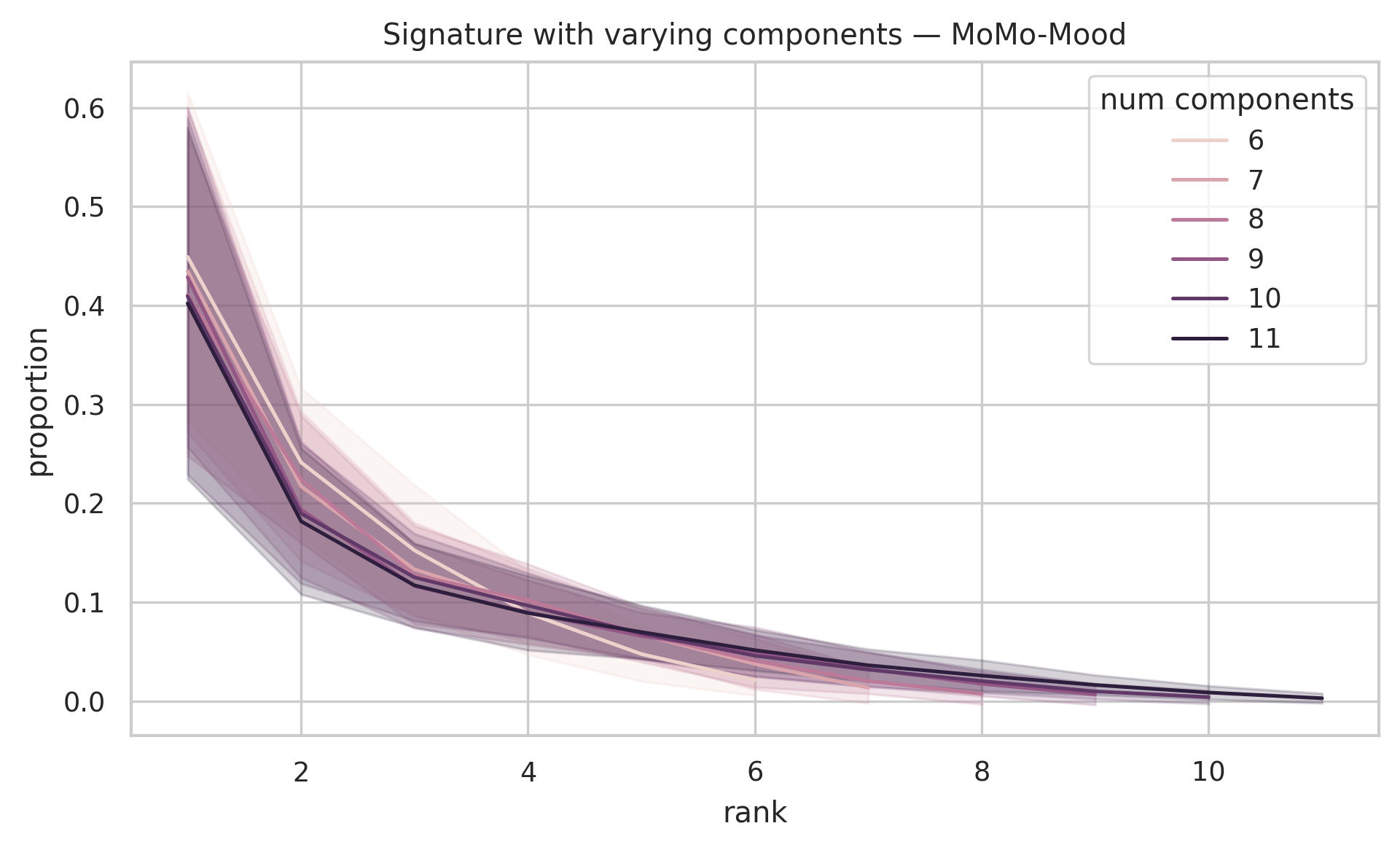}
  \caption{MoMo-Mood. The top routine accounts for a large share, followed by fast decay and a sparse tail, owing to the small sample size. Bands show \(\pm 1\) SD across participants.}
  \label{fig:signature_varyingK_momo}
\end{figure}

\begin{figure}[htbp]
  \centering
  \includegraphics[width=\linewidth]{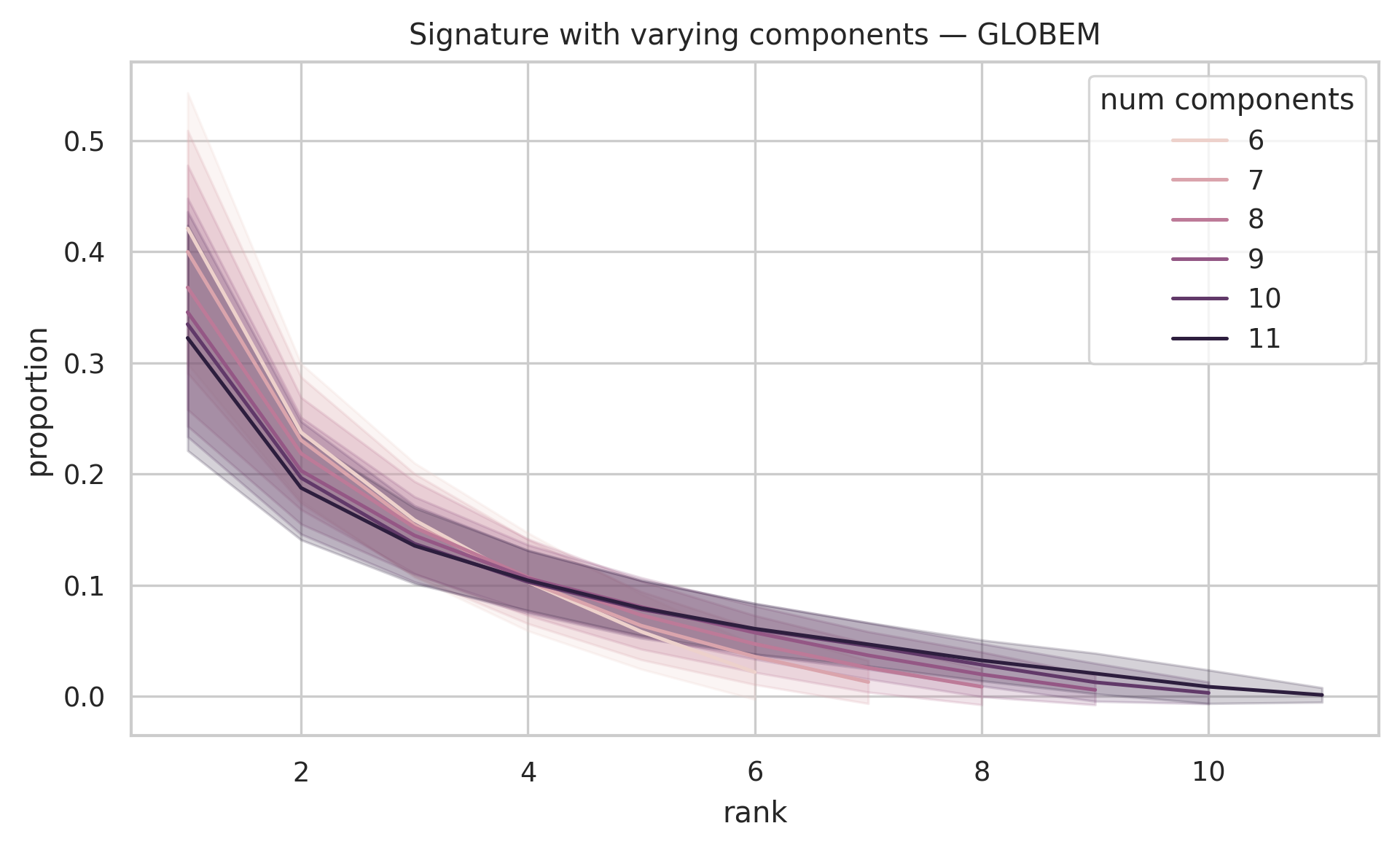}
  \caption{GLOBEM. Same shape as Tesserae: steep head, light tail, and overlap across \(K\) at ranks 1–3; most changes occur in the tail. Bands show \(\pm 1\) SD across participants.}
  \label{fig:signature_varyingK_globem}
\end{figure}

\end{appendices}

\end{document}